\documentclass[letterpaper]{article} 
\usepackage{aaai24}  
\usepackage{times}  
\usepackage{helvet}  
\usepackage{courier}  
\usepackage[hyphens]{url}  
\usepackage{graphicx} 
\usepackage{natbib}  
\usepackage{caption} 
\usepackage{algorithm}
\usepackage{algorithmic}

\usepackage{newfloat}
\usepackage{listings}
\DeclareCaptionStyle{ruled}{labelfont=normalfont,labelsep=colon,strut=off} 
\floatstyle{ruled}
\newfloat{listing}{tb}{lst}{}
\floatname{listing}{Listing}
\usepackage{amssymb}
\usepackage{pifont}
\usepackage{threeparttable}
\usepackage{booktabs}
\usepackage{siunitx}
\usepackage{amsmath}
\usepackage{color-edits}

\addauthor[Aleksandra]{ak}{green}

\addauthor[Jane]{jc}{blue}

\title{Why am I Still Seeing This: Measuring the Effectiveness Of Ad Controls and Explanations in AI-Mediated Ad Targeting Systems}
\author {
    Jane Castleman,
    Aleksandra Korolova
}
\affiliations {
    Princeton University\\
    janeec@princeton.edu, korolova@princeton.edu
}

\begin{document}

\maketitle

\begin{abstract}
Recently, Meta has shifted towards AI-mediated ad targeting mechanisms that do not require advertisers to provide detailed targeting criteria. 
The shift is likely driven by excitement over AI capabilities as well as the need to address new data privacy policies and targeting changes agreed upon in civil rights settlements.
At the same time, in response to growing public concern about the harms of targeted advertising, Meta has touted their ad preference controls as an effective mechanism for users to exert control over the advertising they see. 
Furthermore, Meta markets their ``Why this ad" targeting explanation as a transparency tool that allows users to understand the reasons for seeing particular ads and inform their actions to control what ads they see in the future. 

Our study evaluates the effectiveness of Meta's ``See less'' ad control, as well as the actionability of ad targeting explanations following the shift to AI-mediated targeting. 
We conduct a large-scale study, randomly assigning participants the intervention of marking ``See less'' to either \textit{Body Weight Control} or \textit{Parenting} topics, and collecting the ads Meta shows to participants and their targeting explanations before and after the intervention. 
We find that utilizing the ``See less'' ad control for the topics we study does not significantly reduce the number of ads shown by Meta on these topics, and that the control is less effective for some users whose demographics are correlated with the topic. 
Furthermore, we find that the majority of ad targeting explanations for local ads made no reference to location-specific targeting criteria, and did not inform users why ads related to the topics they requested to ``See less" of continued to be delivered. 
We hypothesize that the poor effectiveness of controls and lack of actionability and comprehensiveness in explanations are the result of the shift to AI-mediated targeting, for which explainability and transparency tools have not yet been developed by Meta.
Our work thus provides evidence for the need of new methods for transparency and user control, suitable and reflective of how the increasingly complex and AI-mediated ad delivery systems operate.
\end{abstract}

\section{Introduction}\label{sec:intro}

In the first quarter of 2024, Meta brought in \$35.6 billion in ad revenue \cite{2024_q1_ad_revenue}, with their targeted advertising and matching systems helping advertisers reach new and existing audiences for whom the ads are particularly relevant~\cite{Meta_2022}. 
However, extensive prior work now details the harms that such targeted ads, pushed to users' feeds based on information collected about them without their explicit control, can have on users. 
For example, ads on subjects such as body weight control and image can increase anxiety and threaten users' health~\cite{Gak_Olojo_Salehi_2022}.
Queer users have pointed out the dangers of targeted advertising in violating their privacy and control over their identity, and expressed the need for better explanations and controls to mediate the information and assumptions used to target their ads \cite{Sampson_Encarnacion_Metaxa_2023}. 
Users with particular health conditions note the helplessness and trauma they feel when ads related to their health are forced onto their feed, with no way to completely stop these ads or prevent targeting based on inferred traits \cite{Wu_Bice_Edwards_Das_2023}. 

\begin{figure}[ht]
    \centering
    \includegraphics[width=\linewidth]{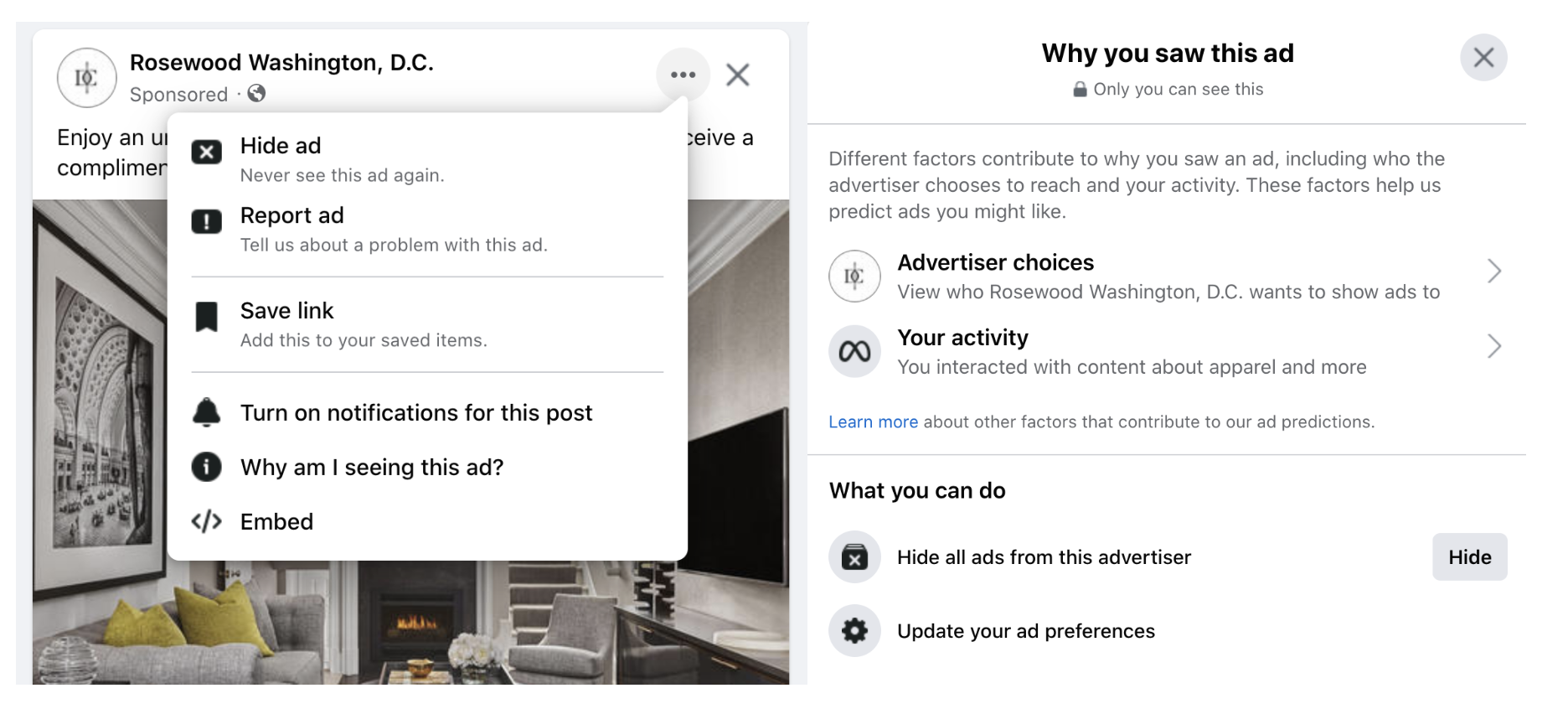}
    \caption{The options presented to users after clicking into the ``Why am I seeing this ad?'' interface.}
    \label{fig:waist}
\end{figure}

In response to these harms, Facebook (currently Meta) built a suite of user-facing tools such as ad preferences and individual ad targeting explanations, aimed to give users control and agency over the ads they see. 
A key \textit{ad personalization control tool} is a button that allows users to mark ``See less'' to specific ad topics in Ad Preferences. Facebook promises that after marking ``See less'' for a topic, advertisers won't be able to target the user by specifying an interest in that topic for all future ads (shown in Figure \ref{fig:ad_responsiveness}) \cite{FacebookSeeLessDescription}. 
A key \textit{transparency control tool} is the ``Why am I seeing this ad?'' interface (shown in Figure \ref{fig:waist}) at the top of each ad~\cite{waist}, which Facebook promises details the advertiser choices and user activity that informed that ad's delivery.
Furthermore, the explanations are aimed to help users inform their ad control choices, as they provide a key entry point to the ad preferences.

Previous work has already found these controls to have limited accessibility \cite{Hsu_Vaccaro_Yue_Rickman_Karahalios_2020} and the ad explanations to not be fully transparent \cite{Andreou_Venkatadri_Goga_Gummadi_Loiseau_Mislove_2018}.
\textbf{Our main focus is understanding whether the effectiveness of these controls and explanations are further hampered by the recent shift to AI-mediated targeting on Meta}. 

The shift to AI-mediated targeting has been a result of the introduction of new privacy protections and restrictions on targeting due to regulatory, competition, civil rights and political considerations. 
In particular, new data privacy policies, such as Apple’s App Tracking Transparency, have pressured Facebook to develop mechanisms to find audiences to deliver ads to while having less direct access to user data, especially from the advertisers. 
Furthermore, in response to a 2018 civil rights settlement on discrimination in employment advertising \cite{ACLU_suit}, a 2022 settlement on discrimination in housing advertising \cite{FacebookvsHUD2}, and continued public pressure on the potential harms of political advertising \cite{Shane_2017, ribeiro2019microtargeting, weintraub2019microtargeting}, Facebook has removed thousands of targeting categories from its advertiser-facing tools from 2022 to 2024~\cite{Facebook_target_2022, Facebook_target_2024, Proxies2024} and now aims to decrease skew in delivery of housing and employment ads \cite{FacebookvsHUD, FacebookvsHUD2023, Bogen2023, Timmaraju_Mashayekhi_Chen_Zeng_Fettes_Cheung_Xiao_Kannadasan_Tripathi_Gahagan_2023}.  

At the same time, to continue delivering on its message of the value of personalized advertising to users and the value of helping the advertisers reach the right audiences, Facebook has introduced new AI-mediated targeting and audience selection tools and new server-side tracking tools \cite{El_Fraihi_Amieur_Rudametkin_Goga_2024}. 
One such AI-mediated audience selection tool, Advantage+ Audiences~\cite{Meta_2022}, promises to improve advertiser experiences and reduce costs \cite{Facebook_target_2020, Facebook_target_2022b, Facebook_target_2024}, using AI to identify relevant users without explicit targeting criteria such as user demographics and interests provided by the advertiser. Instead, Advantage+ relies on data about users' past conversions and browsing behaviors, and ad interactions of users that are algorithmically deemed similar, to narrow audiences on the advertisers' behalf~\cite{Facebook_Advantage+_2024}.

Much of recent AI-based technology suffers from lack of interpretability and explainability, both due to its algorithmic complexity and due to the complexity of the data it relies upon \cite{Lipton_2018, Poursabzi_Sangdeh_Goldstein_2021}. 
We thus hypothesize that the shift to AI-mediated targeting brings new challenges to the effectiveness of controls and explanations related to advertising personalization, as those tools likely have not kept pace with the shift to AI and are likely still built around categorical (i.e. demographic-, interest-, and activity-based) audience creation practices.
We hypothesize that since Meta's suite of new Advantage+ tools no longer requires advertisers to supply precise targeting criteria, there is no clear pathway for users to see particular interests in their ``Why this ad?" explanations, or for the ad delivery algorithm to act upon the ``See less" user actions for specific topics.
The result is lack of effectiveness of ad controls and explanations.

\subsubsection{Our work and contributions}
We perform a large-scale sociotechnical audit to measure how Facebook's ad matching algorithm responds to user-initiated changes in ad controls on the topics of \textit{Body Weight Control} and \textit{Parenting} in early 2024.  
We recruit a generic group of users and a group of users whose age and gender are correlated with interest in each topic to understand whether it is more difficult for users belonging to certain demographic groups to effectively opt-out of certain ad topics. 
We randomly assign our recruited participants to mark ``See less'' to the ad topics and measure whether there is an observable change in the number of ads those users subsequently receive about those topics.
Finally, to provide insight into the impact of Advantage+ audiences on ad explanations, we collect the targeting explanations attached to each ad and quantify the misalignment between the ad content and explanations.

Our findings provide new quantitative evidence of the ineffectiveness of marking ``See less'' to the  topics of \textit{Body Weight Control} and \textit{Parenting}. They also show that ad targeting explanations do not productively inform ad control changes, including for ads relating to topics that users have marked to ``See less'' of. We summarize our findings as follows:

\begin{itemize}
    \item We find large-scale, quantitative evidence of the ineffectiveness of Facebook's ``See less'' ad control that was built around categorical targeting of demographics, interests, and behaviors.
    \item We compare the effectiveness of ``See less'' ad controls across demographics, and find the controls are less effective for some users belonging to demographic groups that are more likely to have actual experiences with the topic they are trying to ``See less'' of. 
    \item In our analysis of ad targeting explanations we find that the majority fail to be informative and do not connect to existing ad controls. In particular, the misalignment of generic geographic targeting explanations with evidently local ad content offers insight into the effects of Advantage+ audience targeting on actionability of ad explanations. 
\end{itemize}

These results validate our hypothesis that AI-mediated targeting mechanisms negatively impact the effectiveness of ad controls and the actionability of ad delivery decisions, leading to a substantial decrease in meaningful transparency and control over advertising personalization for the users.
Our work thus motivates new research questions in explainability of AI, particularly in reconciling AI-mediated targeting and matching with the need for faithful, actionable explanations and controls.

\section{Related Work}\label{sec:related_work}

Existing research outlines the importance of user control, transparency, and algorithmic audits in supporting users and holding platforms accountable for their targeted advertising systems.

\subsubsection{User Control}

Providing users with control over the ads they see is crucial for maintaining their agency and safety. Ad delivery algorithms continue to show clickbait, untrustworthy, and distasteful ads despite including ad quality in their ad ranking process \cite{Zeng_Kohno_Roesner_2021}. Ads can also cause psychological distress, loss of autonomy, changes to user behavior, and marginalization or traumatization \cite{Wu_Bice_Edwards_Das_2023}.

Our study of Facebook's ad controls focuses on ad topic controls, inferred from users' activities and used in ad explanations. For users with histories of disordered eating, targeted advertising worsened anxiety surrounding food and exercise and negatively impacted self-esteem \cite{Gak_Olojo_Salehi_2022}. For years, women have noted their inability to see fewer ads related to pregnancy and children's health, increasing anxiety and forcing them to revisit trauma \cite{Brockell_2018, Contreras_2022}. While users cannot fully remove assigned ad topics, they can mark ``See less'' to ads relating to these topics. Given users' expressed desires to exert control over ads relating to \textit{Body Weight Control} and \textit{Parenting}, we focus on these topics in our study of effectiveness of Facebook's ``See less'' ad control. 

Despite Facebook's communications regarding the power and value of its ad personalization and transparency control tools \cite{waist, FacebookSeeLessDescription}, research continues to find that current systems have ineffective algorithmic controls, opaque and misleading explanations, and poorly designed interfaces, all reducing user agency \cite{Chromik_Eiband_Völkel_Buschek_2019}. Other studies have emphasized the lack of usability of existing ad interfaces, \cite{Habib_Pearman_Wang_Zou_Acquisti_Cranor_Sadeh_Schaub_2020, Habib_Pearman_Young_Saxena_Zhang_Cranor_2022, Leon_Ur_Shay_Wang_Balebako_Cranor_2012}, arguing that controls are not sufficiently accessible nor are they aligned with user needs.

\subsubsection{Algorithmic Transparency}

Given the complexity of AI-mediated ad targeting systems, algorithmic transparency is crucial for users' understanding of how their data is used to make ad delivery decisions and inform ad control changes. When users do not fully understand AI systems, they build their own inferences about how the algorithm functions \cite{Yuan_Bi_Lin_Tseng_2023} and expect improvement \cite{Smith-Renner_Fan_Birchfield_Wu_Boyd-Graber_Weld_Findlater_2020}, which can lead to misinformed attempts at control and frustration. Users also note the importance of accurate ad explanations for ads leveraging personal data, interest targeting, and custom audiences to maintain a sense of control over their data \cite{Lee_Logas_Yang_Li_Barbosa_Wang_Das_2023, Wei_Stamos_Veys_Reitinger_Goodman_Herman_Filipczuk_Weinshel_Mazurek_Ur_2020}, and their desire to understand how algorithms use their information to make inferences \cite{Dolin_Weinshel_Shan_Hahn_Choi_Mazurek_Ur_2018, Eslami_Krishna_Kumaran_Sandvig_Karahalios_2018}. Transparent explanations and opportunities for feedback through ad controls help inform productive choices over AI-mediated decisions and user data \cite{Smith-Renner_Fan_Birchfield_Wu_Boyd-Graber_Weld_Findlater_2020}. 

Recent research by \cite{Chouaki_Bouzenia_Goga_Roussillon_2022} suggests that changes to Facebook's ad platform have driven a shift away from advertiser-driven microtargeting, finding that the majority of ads did not use microtargeting in comparison to a 2018 study \cite{Cabanas_Cuevas_Cuevas}. 
We hypothesize that shift towards generic, AI-mediated targeting threatens transparency, since ad delivery algorithms can no longer rely on advertiser targeting criteria to explain ad delivery, creating new transparency challenges. 

Additionally, many problematic ads use no targeting, indicating that the ad delivery algorithm perceives them as relevant to users rather than delivering them based on audience targeting \cite{Ali_Goetzen_Mislove_Redmiles_Sapiezynski_2023}. When ad explanations are incomplete or incorrect, users may still attempt to infer explanations and are hesitant to blame algorithmic error, misinforming future actions to control their ads \cite{Eslami_Krishna_Kumaran_Sandvig_Karahalios_2018, Rader_Gray_2015}. Complete, accurate, and actionable ad explanations are thus necessary for expressing what data was used to inform ad delivery and provide users with information to make productive ad control changes.

\subsubsection{Algorithmic Audits}

Algorithmic audits are essential tools for investigating the mechanisms of black-box ad delivery systems to uncover their impacts on privacy, fairness, polarization, and user experiences. Previous investigations into Facebook’s ad targeting revealed that it led to privacy threats to users' personal information~\cite{faizullabhoy2018facebook, korolova2011privacy}; the Meta Pixel sent sensitive medical information to advertisers \cite{Feathers_Fondrie_Mattu_2022}, and that adversaries could uncover users’ phone numbers and site visits by reconstructing custom audience data \cite{Venkatadri_Andreou_Liu_Mislove_Gummadi_Loiseau_Goga_2018}. 
Audits also revealed discriminatory ad delivery, with algorithms inequitably delivering harmful ads \cite{Ali_Goetzen_Mislove_Redmiles_Sapiezynski_2022}, housing and employment ads \cite{Ali_Sapiezynski_Bogen_Korolova_Mislove_Rieke_2019,  Nagaraj_Rao_Korolova_2023, imana2021auditing}, and education ads \cite{Imana_Korolova_Heidemann_2024}, and contributing to echo chambers in political advertising~\cite{ali2021ad}. Researchers have also found that custom audiences can introduce bias \cite{Sapiezynski_Ghosh_Kaplan_Rieke_Mislove_2022} and amplify existing bias \cite{Speicher_Ali_Venkatadri_Ribeiro_Arvanitakis_Benevenuto_Gummadi_Loiseau_Mislove} in advertiser audiences. Large-scale user audits provide insight into user experiences with ad systems, such as issues with ad controls \cite{Datta_Tschantz_Datta_2015, Habib_Pearman_Young_Saxena_Zhang_Cranor_2022}, perceptions of problematic ads \cite{Zeng_Kohno_Roesner_2021}, and preferred ad explanations \cite{Lee_Logas_Yang_Li_Barbosa_Wang_Das_2023, Wei_Stamos_Veys_Reitinger_Goodman_Herman_Filipczuk_Weinshel_Mazurek_Ur_2020}.

Using a large-scale user audit, our study addresses a gap in existing research by providing quantitative evidence of the ineffectiveness of Facebook's ad topic controls and the poor actionability of their targeting mechanism, prompted by the shift to AI-mediated targeting, marketed as Advantage+ audiences.

\section{Facebook's Commitments to Users}

\subsection{Ad Controls}

\begin{figure}[ht]
    \centering
    \includegraphics[width=\linewidth]{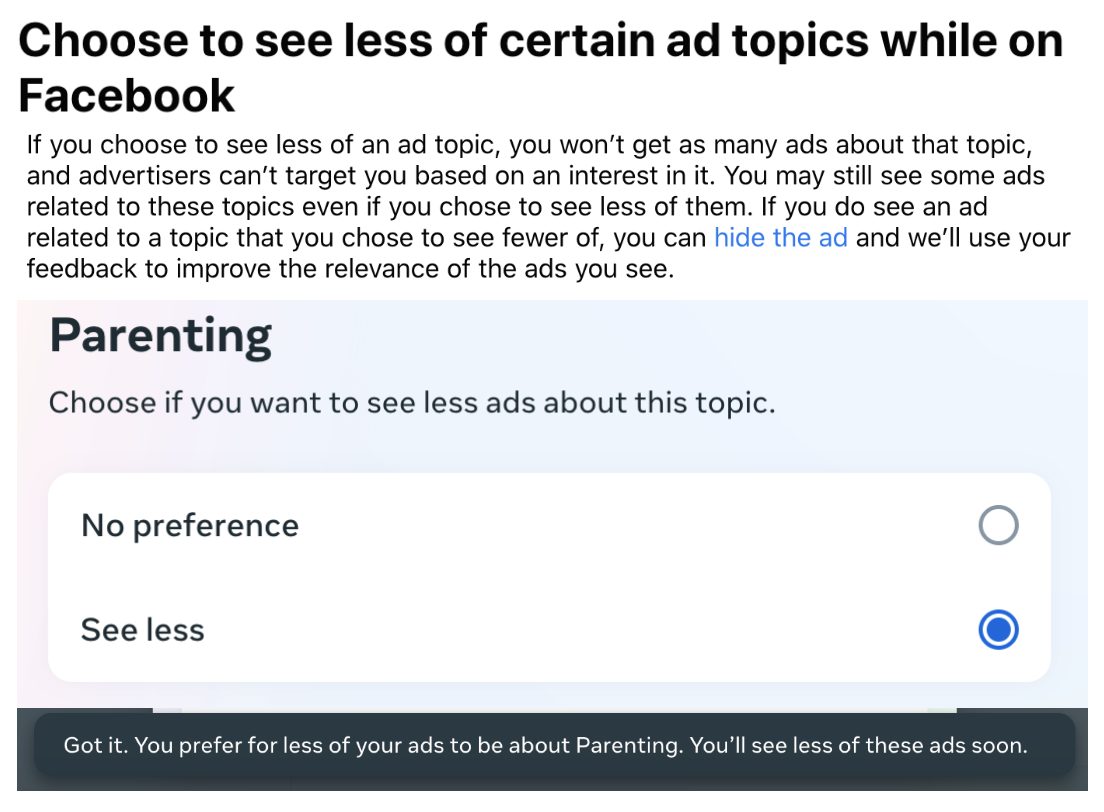}
    \caption{Facebook's description of the outcome of marking ``See less'' \cite{FacebookSeeLessDescription}, and the message returned when the ``See less'' control is changed for the topic \textit{Parenting}.}
    \label{fig:ad_responsiveness}
\end{figure}

Facebook introduced the ``See less'' ad control in 2021 followed by an update in 2022, just four months before the release of Advantage+ audiences \cite{Meta_2022}, promising that the ad topic control allows users to restrict the interest targeting categories used to reach them and the exert control over the ad content they see by opting to see fewer ads relating to certain topics \cite{Facebook_target_2022}. Figure \ref{fig:ad_responsiveness} shows their claim to users that ``you won't get as many ads about that topic and advertisers can't target you based on an interest in it'' \cite{FacebookSeeLessDescription}. 

When we tested the ``See less'' control in 2024, Facebook returned a popup message indicating that an ad control setting had been changed, shown in Figure \ref{fig:ad_responsiveness}. The commitment to show fewer related ads ``soon'' is vague, but we still expect users to see fewer related ads to a topic they have marked to ``See less'' of over a period of a couple weeks.

\subsection{Ad Explanations} 
One of Facebook's five pillars of AI \cite{Pesenti_2021} is ``Transparency \& Control.'' One mechanism to uphold this pillar is the ``Why am I seeing this ad?'' interface attached to each ad, initially debuted in 2014 \cite{waist}. The ``Why am I seeing this ad?'' interface includes an ``Advertiser Choices'' section listing an advertiser's targeting mechanisms, such as profile information, custom audiences, interests, and location \cite{waist}. A 2023 update adds a ``Your activity'' section listing previous user activity that influenced ad delivery, outputted by machine learning models similar to the ones that inform ad delivery (shown in Figure \ref{fig:targeting_coding}) to further increase transparency \cite{Meta_Transparency_2023}. To be fully transparent, we expect these explanations to contain all of the ``Advertiser Choices'' that led to a user being targeted.

\section{Methodology}\label{sec:method}

In this section, we describe the structure of our user survey to perform a large-scale sociotechnical audit of Facebook's ad controls and ad explanations for two topics: \textit{Parenting} and \textit{Body Weight Control}. Our goal is to evaluate whether they uphold Facebook's commitments to users made in their descriptions of ad explanations and ad controls.

\subsection{Participant Recruitment}
We recruit participants for our study on Prolific \cite{prolific} and use Prolific's built-in screeners based on participants' self-identified characteristics to choose two sets of Prolific users: one set with users who may be more likely to have ads related to \textit{Parenting} and one set with users who may be more likely to have ads related to \textit{Body Weight Control}. Given the inherently limited scope of our study, we chose these topics due to previously expressed concerns about ads worsening trauma related to parenting \cite{Brockell_2018, Contreras_2022} or exacerbating food- and exercise-related anxiety \cite{Gak_Olojo_Salehi_2022}.
Participants in each set match the following screening characteristics:

\begin{itemize}
    \item \textit{Parenting}: Has a child less than 10 years old, currently living with their child, uses Facebook.
    \item \textit{Body Weight Control}: Has gone on a diet in the past, marked ``Health \& Fitness'' as a hobby, exercises ``Sometimes'' or ``Often,'' uses Facebook.
\end{itemize}

Furthermore, to test our hypothesis that it is more difficult for users belonging to demographic groups that Facebook deems correlated with the topics to effectively change the ads they see via the provided ``See less" controls we develop the following \textit{correlated demographics} for each interest, used in combination with our previous screeners:  

\begin{itemize}
    \item \textit{Parenting}: Women aged 25 to 45, since they are most likely to be new parents \cite{Bui_Miller_2018}.
    \item \textit{Body Weight Control}: Women aged 18 to 59 and men aged 40 to 59, since these groups are most likely to have experiences with dieting and weight loss \cite{Martin_Herrick_Sarafrazi_Ogden_2019}.
\end{itemize}

In sum, we recruit 201 participants, 110 from the \textit{Parenting} screener, of whom 68 match the \textit{Parenting} correlated demographic, and 91 from the \textit{Body Weight Control (BWC)} screener, of whom 31 match the correlated demographic. Table ~\ref{table:demographics} shows the demographic breakdown of our participants. Our participants' demographics are skewed towards women and individuals under 50, in line with the skewed demographics of Prolific users \cite{prolific_demographics}.

\begin{table}[b!]
\centering
\begin{tabular}{
  @{}
  l
  l
  S[table-format=2.1]
  S[table-format=2.1]
  S[table-format=2.1]
  S[table-format=2.1]
  @{}
}
\toprule
\textbf{Variable} & \textbf{Value} & \multicolumn{2}{c}{\textit{\textbf{Parenting}}} & \multicolumn{2}{c}{\textit{\textbf{ BWC}}} \\

                  &                & {\textbf{n}} & {\textbf{\%}}            & {\textbf{n}} & {\textbf{\%}}        \\
\midrule
\textbf{Gender} & Woman & 92 & 83.6 & 57 & 63.3 \\
        & Man & 18 & 16.4 & 34  & 36.7  \\
        \\
\textbf{Age}   & 18-29 & 23   & 22.1  & 17   & 18.7  \\
               & 30-49 & 70  & 67.3 & 46 & 50.6  \\
               & 50-69 & 12  & 11.5 & 25 & 27.5  \\
               & $> 69$ & 0  & 0 & 3 & 3.3  \\
\\
\textbf{Correlated}  & 1 (yes) & 68  & 61.8  & 31  & 34.1 \\
                  & 0 (no) & 42 & 38.2 & 60 & 65.9 \\
\midrule
Total             &                & 110         &                          & 91          &                      \\
\bottomrule
\end{tabular}
\caption{Demographics of participants, stratified by topic.}
\label{table:demographics}
\end{table}

\subsection{Study Design}

\begin{figure*}
\centering
\includegraphics[width=0.9\linewidth]{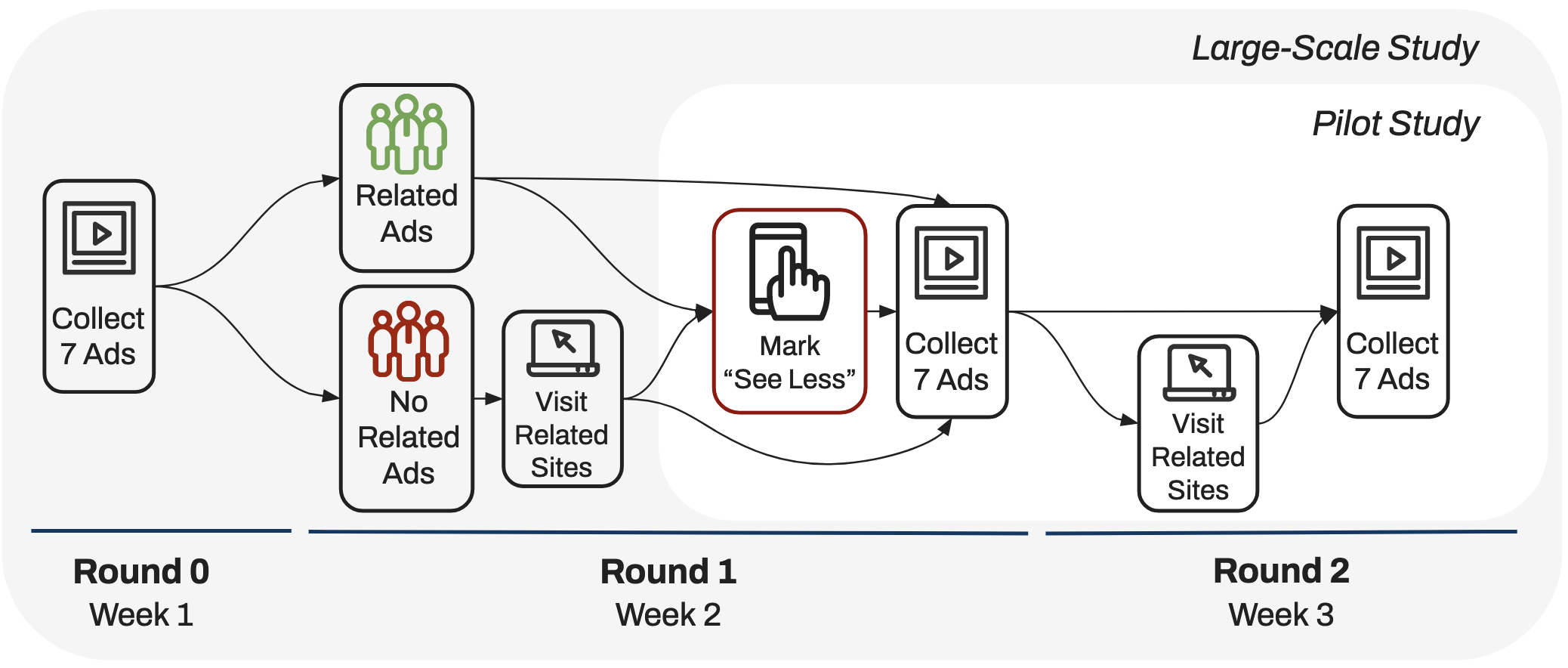}
\caption{Timeline of our user survey to collect feed ads with randomized ad control intervention.}
\label{fig:method}
\end{figure*}

We asked all participants to collect the ads they see on Facebook and share them in three separate rounds of the study, separated by approximately one week each, in a process outlined in Figure \ref{fig:method}. 

After analyzing the set of shared ads from Round 0, we asked the 62 participants for the topic \textit{Parenting} and 79 participants for the topic \textit{Body Weight Control} who did not see any ads related to their assigned topic to visit the websites of advertisers related to their topic. 
During Round 1, we asked a randomly chosen half of participants from each topic group, whom we subsequently call our intervention group, to mark ``See less'' to their assigned topic, then share the ads they saw immediately after that. For the topic \textit{Parenting}, 46 participants exercised the ``See less'' ad control and for the topic \textit{Body Weight Control}, 28 participants exercised the ``See less'' ad control. The other participants, whom we subsequently call our Control group, were not asked to perform any action and were only asked to share the ads they saw, as in Week 1. During Round 2, we again asked the groups of participants who did not see any ads related to their assigned topic to visit the websites of advertisers related to their topic.

We selected the websites for participants without related ads to visit from stereotypical advertisers in \textit{Parenting} and \textit{Body Weight Control}, prioritizing selection using the number of advertisements the advertiser was actively running as listed by the Facebook Ad Library. Then, we verified their websites use Meta Pixels to track participant data using The Markup's BlackLight tool \cite{Mattu_Sankin_2020}. We did this to simulate users with continued web activity to sites related to topics they marked to ``See less'' of, which should not reduce the effectiveness of the ``See less'' ad control. 

For each ad, we collected the advertiser, the ``Why am I seeing this ad?'' information attached to each ad, and a link to the ad. The large-scale study consisted of one pre-intervention round (Round 0), collecting 8 feed ads per participant, followed by 2 post-intervention rounds (Round 1 and Round 2), collecting 7 feed ads per participant. The pilot study consisted of Round 1 and Round 2 from the large-scale study, including the randomized intervention and the sorting of participants into groups with related and unrelated ads.

\begin{table}[ht]
\centering
\begin{tabular}{ll}
\hline
Parenting & Body Weight Control \\ \hline
Goodnites & Planet Fitness \\
Primrose Schools & Noom \\
Care.com & WeightWatchers \\
Pampers & Nutrisystem \\
Graco Baby  & OrangeTheory \\
Kindercare & WHOOP \\
BuyBuyBaby & MyFitnessPal \\
The Nok Box &  \\
\hline
\end{tabular}
\caption{Advertisers related to participants' topic of interest.}
\label{table:sites}
\end{table}

\section{Measuring Ad Control Effectiveness}

We first study the effectiveness of the ``See less'' control by measuring how the average number of ads related to topic marked ``See less'' changes over time for the intervention versus control groups. We split our data into two datasets, one for participants assigned the topic \textit{Parenting} and one for \textit{Body Weight Control}. Then, we create a codebook to assign each ad a score based on its relatedness to a participant's assigned topic, calculating the average relatedness for the intervention and control groups in each dataset. We find that the intervention and control groups experience similar rates of related ads over time, suggesting that the ``See less'' intervention may not be effective.

\subsection{Classifying Related Ads}

We manually code the relatedness of each ad to the participant's assigned topic, either \textit{Parenting} or \textit{Body Weight Control}, along the following scale: “Not related” = 0, “Somewhat related” = 0.5” and “Related” = 1. We assign each ad $j$ with a numeric $ad\_score[j]$ using this relatedness scale.\footnote{The statistical significance of our results is robust to changes in the magnitude of our ad relatedness scale; values from 0 to 1 were chosen for simplicity.}

\begin{itemize}
    \item  Score of 0: The text or image does not contain information related to the blocked topic. 
    \item Score of 0.5: The text or image contains information somewhat related to the blocked topic. \textit{Body Weight Control}: diet-focused meal planning services, exercise equipment, and ad images referencing fitness and nutrition. \textit{Parenting}: children's activities, toys, and images of children with references to parenting. 
    \item Score of 1: The text or image contains information directly related to the blocked topic. \textit{Body Weight Control}: fitness products and services, diet plans, and ad images promoting weight loss and working out. \textit{Parenting}: childcare services, childcare products, and children's healthcare.
\end{itemize}

\begin{figure}[tb]
\centering
\includegraphics[width=\linewidth]{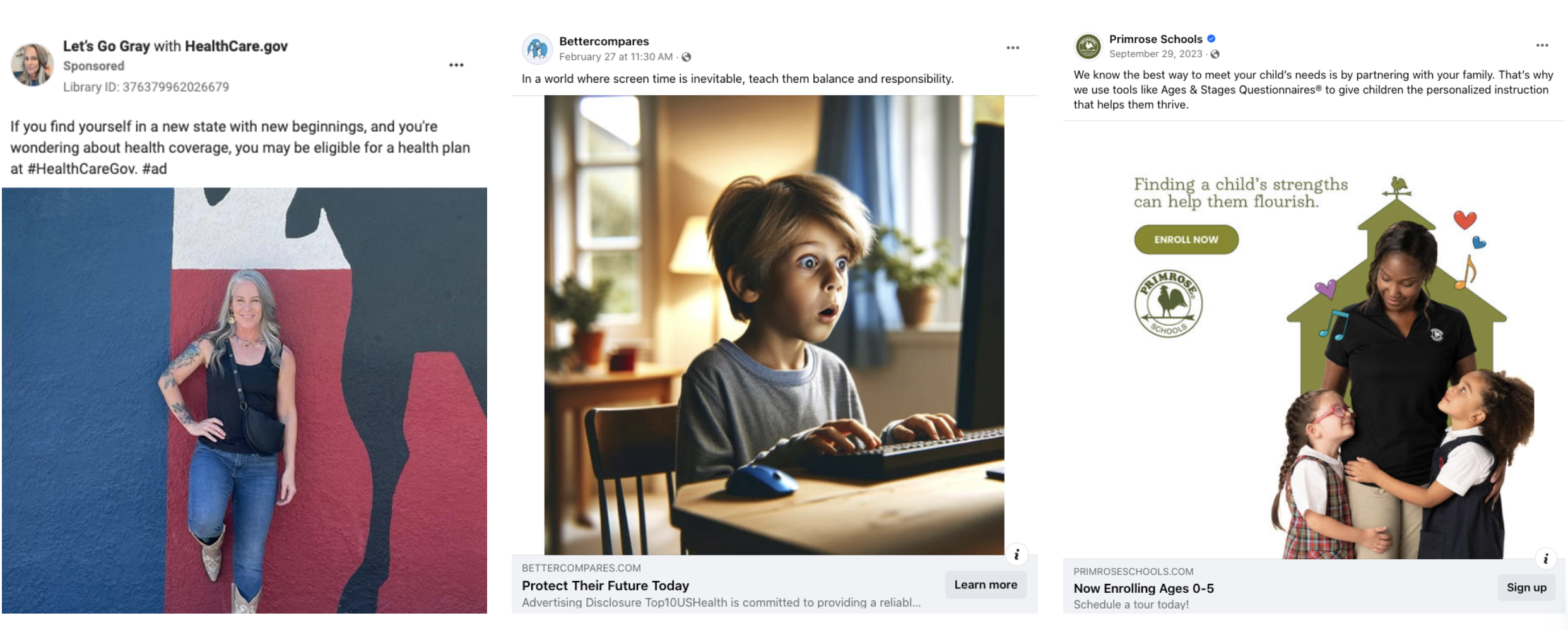}
\caption{From left to right, ads coded with relatedness = 0, 0.5, 1 to \textit{Parenting}.}
\label{fig:parenting_coding}
\end{figure}

\begin{figure}[tb]
\centering
\includegraphics[width=\linewidth]{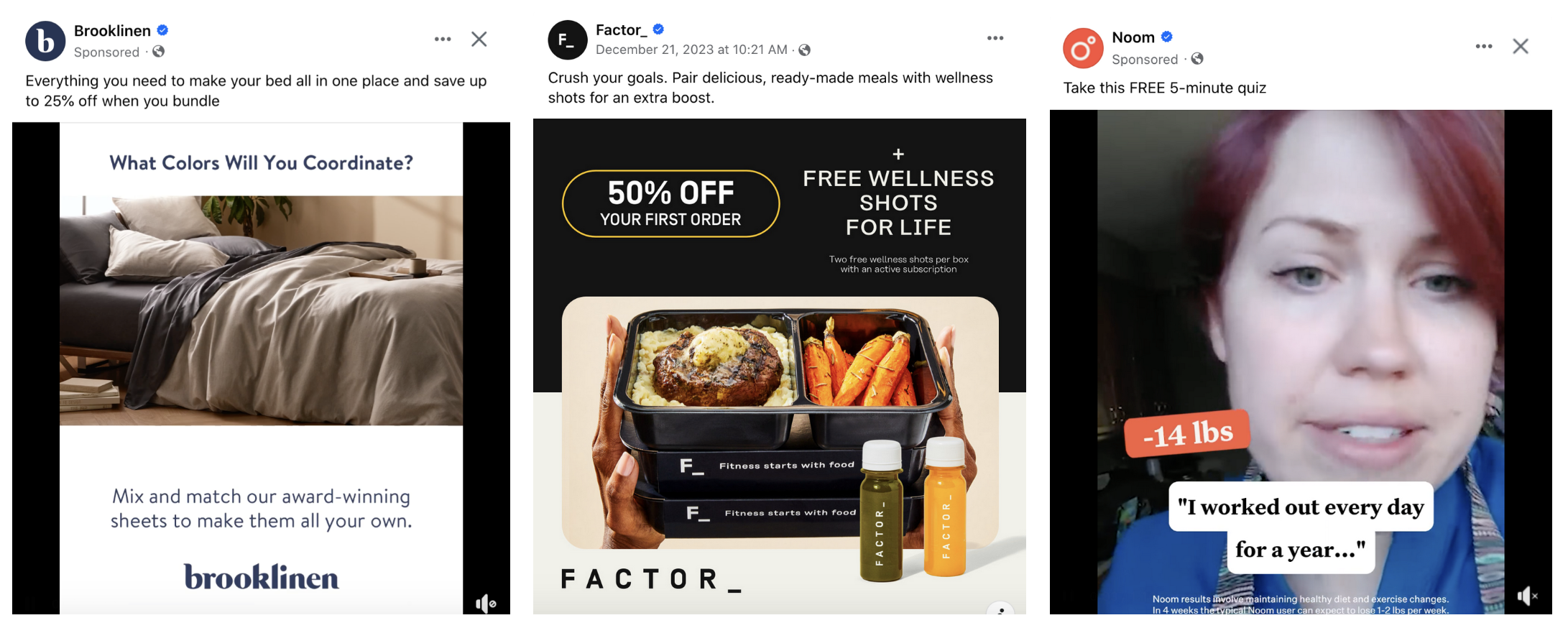}
\caption{From left to right, ads coded with relatedness = 0, 0.5, 1 to \textit{Body Weight Control}.}
\label{fig:bwc_coding}
\end{figure}

Example ads for \textit{Parenting} and \textit{Body Weight Control} are shown in Figures \ref{fig:parenting_coding} and \ref{fig:bwc_coding}, respectively. 

\subsection{Measuring Change in Related Ads}

For each topic, we combine participant data from the pilot and large-scale studies by round. We calculate $r^{(i)}_u$, the proportion of ads related to their assigned topic shown to a user $u$ in Round $i$, by summing the ad scores for each ad $j$ they encountered in Round $i$, and dividing that sum by the total number of ads for that user in that round, $n_{u_i}$:

\begin{equation*}
r^{(i)}_u = \frac{1}{n_{u_i}} \sum^{n_{u_i}}_{j = 1} ad\_score[j]
\label{eq:prop}
\end{equation*}

For a subset of participants $S$ belonging to a particular demographic or intervention group, we calculate the average proportion of related ads in each round. 

\begin{equation*}
\mu_{S}^{r^{(i)}} = \frac{1}{|S|} \sum_{u \in S} r^{(i)}_u
\label{eq:mean_avg}
\end{equation*}

We present the results broken down by round, topic and group in Table \ref{tab:mu_combined_wide}.
For the topic \textit{Parenting}, the control group sees an increase in the proportion of related ads from Round 0 to Round 2, while the intervention group sees a decrease overall, but a slight increase from Round 1 to Round 2. 
However, for the \textit{Body Weight Control} group, both the control and intervention groups see an increase in the proportion of related ads, suggesting the ``See less'' ad control is ineffective.

\begin{table}[tb]
\centering
\begin{tabular}{lcccc}
\hline
 & \multicolumn{2}{c}{\textbf{Parenting}} & \multicolumn{2}{c}{\textbf{Body Weight Control}} \\
\textbf{Round} & Control & Intv. & Control & Intv. \\ 
\hline
0 & 0.130 & 0.163 & 0.029 & 0.038 \\
1 & 0.123 & 0.108 & 0.120 & 0.060 \\
2 & 0.191 & 0.139 & 0.161 & 0.114 \\  
\hline
\end{tabular}
\caption{Average proportion of ads $\mu_{S}^{r^{(i)}}$ related to the topics \textit{Parenting} and \textit{Body Weight Control} seen by users with and without marking ``See less.''}
\label{tab:mu_combined_wide}
\end{table}

\subsection{Testing for Significance in the Change in Relatedness}

To establish whether the observed change in the proportion of related ads over time is statistically significant, we use the Mann-Whitney $U$ test with the significance level $\alpha = 0.1$ \cite{Mann_Whitney_1947}. Specifically, we compare the median change in related ads over time between the control and intervention groups. Our analysis did not show a statistically significant difference in the decrease of related ads between the control and intervention groups for \textit{Parenting} or \textit{Body Weight Control}, suggesting that the ``See less'' ad control does not meaningfully reduce the number of ads related to the blocked topic.

First, we calculate the change in relatedness from Round 0 to Round 2 for user $u$ as $\delta_u$:

\begin{equation*}
\delta_u = r^{(2)}_u - r^{(0)}_u.
\label{eq:delta_prop}
\end{equation*}

Then, we find the median change in the proportion of related ads experienced by a subset of participants in group $S$, $m_\delta^S$:

\begin{equation*}
m_\delta^S = median(\{\delta_u\}, \text{for } u \in S).
\label{eq:median_s}
\end{equation*}

We compare $m_\delta$ for the control and intervention groups, denoted $m_\delta^{P, C}, m_\delta^{P, I}$ for the topic \textit{Parenting} and $m_\delta^{B, C}, m_\delta^{B, I}$ for the topic \textit{Body Weight Control}. Our null hypotheses are $m_\delta^{P, C} = m_\delta^{P, I}$ and $m_\delta^{B, C} = m_\delta^{B, I}$. Table \ref{tab:pvalues_bh} lists the results of these tests.

As the $p$-values for both hypotheses are greater than the significance level of $\alpha = 0.1$, we conclude that there is no evidence that the intervention group saw a significantly greater decrease in the proportion of related ads from Round 0 to Round 2 in comparison to the control group for both \textit{Parenting} and \textit{Body Weight Control}.

\begin{table}[ht]
\centering
\begin{threeparttable}
\begin{tabular}{lll}
\hline
\textbf{Null Hypothesis} & \textbf{$p$-value} & \textbf{Adjusted $p$-value} \\
\hline
\\
$m_\delta^{B, C} = m_\delta^{B, I}$ & $0.129$ & $0.258 $  \\
\\
$m_\delta^{P, C} = m_\delta^{P, I}$ & $0.229$ & $0.305$  \\
\hline
\end{tabular}
\begin{tablenotes}
\small
\item $^{***} p < 0.01, ^{**} p < 0.05, ^{*} p < 0.1$
\end{tablenotes}
\end{threeparttable}
\caption{$p$-values for Mann-Whitney $U$ Tests of Intervention vs. Control.}
\label{tab:pvalues_bh}
\end{table}

\subsection{Comparison of Demographic Type} 

We compare the effectiveness of the ``See Less'' intervention for our topics of interest for all users in the intervention group, split into correlated versus non-correlated demographics. We hypothesize that the effectiveness of the intervention decreases for participants belonging to demographics correlated with our topic of interest. Table \ref{tab:mu_correlated} shows the average proportion in related ads seen by round and demographic type, split by topic. 

The impact of participant demographic on the average proportion of related ads shown after the intervention is mixed for the topics \textit{Body Weight Control} and \textit{Parenting}.
We find that for the topic \textit{Body Weight Control}, participants in non-correlated demographics see a stronger increase in the average proportion of related ads shown than participants in correlated demographics, while for the topic \textit{Parenting}, participants in correlated demographics see an increase in the average proportion of related ads shown while those in non-correlated demographics see a decrease. 

\begin{table}[ht]
\centering
\begin{tabular}{lcccc}
\hline
 & \multicolumn{2}{c}{\textbf{Parenting}} & \multicolumn{2}{c}{\textbf{Body Weight Control}} \\
\textbf{Round} & Corr. & Not Corr. & Corr. & Not Corr. \\ 
\hline
0 & 0.155 & 0.187 & 0.051 & 0.010 \\
1 & 0.132 & 0.048 & 0.072 & 0.028 \\
2 & 0.183 & 0.041 & 0.093 & 0.214 \\  
\hline
\end{tabular}
\caption{Average proportion of ads related to the topics \textit{Parenting} and \textit{Body Weight Control} seen by users in correlated demographics and those not in correlated demographic.}
\label{tab:mu_correlated}
\end{table}

\subsubsection{Testing for Significance across Demographic Types}

We also compare the median change in related ads over time between participants who marked ``See less'' belonging to correlated demographics versus those not belonging to correlated demographics using a Mann-Whitney $U$ test. We compare $m_\delta$ for correlated and non-correlated demographics, denoted $m_\delta^{P, COR}, m_\delta^{P, NCOR}$ for the topic \textit{Parenting} and $m_\delta^{B, COR}, m_\delta^{B, NCOR}$ for the topic \textit{Body Weight Control}. Our null hypotheses are $m_\delta^{P, COR} = m_\delta^{P, NCOR}$ and $m_\delta^{B, COR} = m_\delta^{B, NCOR}$. 
Table \ref{tab:pvalues_cor} lists the results of this test. 

\begin{table}[ht]
\centering
\begin{threeparttable}
\begin{tabular}{llll}
\hline
\textbf{Null Hypothesis} &  \textbf{$p$-value} & \textbf{Adjusted $p$-value} \\
\hline
$m_\delta^{B, COR} = m_\delta^{B, NCOR}$ & $0.995$ & $0.995 $  \\
$m_\delta^{P, COR} = m_\delta^{P, NCOR}$ &  $0.021^{**}$ & $0.085^*$  \\
\hline
\end{tabular}
\begin{tablenotes}
\small
\item $^{***} p < 0.01, ^{**} p < 0.05, ^{*} p < 0.1$
\end{tablenotes}
\end{threeparttable}
\caption{$p$-values for Mann-Whitney $U$ Tests between Demographic Types.}
\label{tab:pvalues_cor}
\end{table}

For the topic \textit{Parenting}, participants in correlated demographics see significantly less of a decrease in related ads than participants in non-correlated demographics. This provides statistically significant evidence at $\alpha = 0.1$ level the that it is more difficult for users in our defined correlated demographics to see fewer ads related to \textit{Parenting} than users not in correlated demographics.
Conversely, we find no significant difference in the median change in the proportion of related ads between participants who marked ``See less'' in correlated versus non-correlated demographics for the topic \textit{Body Weight Control}.

\section{Measuring Actionability of Targeting Explanations}
Facebook does not publish information about the number or types of advertisers using the suite of Advantage+ products, nor is this information included in the ``Why am I seeing this?'' interface. 
Therefore, to gain circumstantial evidence of the prevalence of Advantage+ audience use, we first measure the frequency of location-specific criteria in the ad targeting information of local ads. 
We then measure whether topic-specific ads indicate that they were targeted using that particular topic. 
We hypothesize that ads delivered using Advantage+ audience selection only return generic information about their location and interest targeting in the explanations, and thus the above measurements would be good proxies.
We find that the majority of location-specific ads do not reference location information in targeting explanations, and the majority of ads related to specific interests only return generic targeting information.
The findings support our hypothesis that the shift to Advantage+ audience selection reduces ad explanation faithfulness and actionability. 

\subsection{Defining Local Ads and Targeting Explanation Types}
We collected 3,868 ads from 201 participants over three rounds. We now detail how we label ads as location-specific and classify targeting information into three types: generic, activity-based, and interest-based.

\subsubsection{Labeling Local Ads} To classify ads as \textit{local}, we establish specific criteria based on the ad creative’s content. An ad is labeled as local if it references products and services that are geographically constrained, meaning they are only accessible to users within a particular area. We use the following criteria to determine whether an ad is local: 

\begin{itemize}
    \item Geographic references: ad explicitly references a specific state, city, town, or neighborhood, excluding vacation ads.
    \item Location-based services: ad references services that are inherently local, such as real estate listings, local events, or health services.
    \item Contact Information: ad lists local phone number, address, or directions indicating a specific location.
\end{itemize}

\subsubsection{Labeling Targeting Explanations}

\begin{figure*}
\centering
\includegraphics[width=\linewidth]{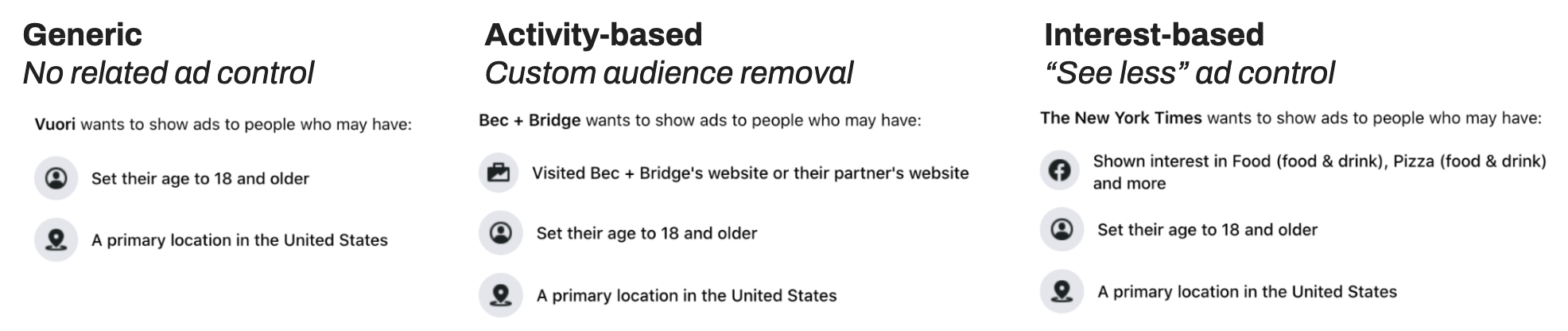}
\caption{From left to right, ``Why am I seeing this ad?'' explanations classified as generic, interest-based, and activity-based.}
\label{fig:targeting_coding}
\end{figure*}

Using the advertising targeting information from the uploaded ``Why am I seeing this ad?'' explanation, we label each targeting explanation as \textit{generic}, \textit{interest-based}, or \textit{activity-based}, with examples shown in Figure \ref{fig:targeting_coding}.

We label ad explanations that only included a location as the United States, age, language, and gender as \textit{generic}, since users cannot change these attributes. Generic explanations do not correspond to an existing ad control, and do not give information as to why the ad was delivered to a specific user since they are so broad. Ads using Advantage+ audience selection return generic ad explanations since they do not rely on manual targeting criteria. 

We label explanations listing an interest as \textit{interest-based}, which connect inferences of users interests to ad delivery and inform changes to the ``See less'' ad control. We label ad explanations listing web activity, including similar activity to existing customers, as \textit{activity-based}, which connect online activity to ad delivery and inform changes to custom audience preferences. Ad explanations can be both interest-based and activity-based. Our full criteria is listed in Table \ref{table:targeting_info}. A ``\checkmark'' indicates included information and a ``\ding{55}'' indicates excluded information. A label is added to an explanation when all included and excluded information is satisfied. 

\begin{table}[tb]
\centering
\begin{tabular}{lccc}
\hline
Targeting Info. & Generic & Activity & Interest \\ \hline
Age & \checkmark & - & -  \\
Location (broad) & \checkmark  & -  & - \\
Language  & \checkmark &  -  & - \\
Gender & \checkmark   &  -  &  - \\
Interest & \ding{55}   &  -  &  \checkmark \\
Site Visit & \ding{55}   &  \checkmark  &  - \\
Hashed List & \ding{55}   &  \checkmark  &  - \\
Lookalike Audience & \ding{55}   &  \checkmark  &  - \\
\hline
\end{tabular}
\caption{Coding Ads by Targeting Information}
\label{table:targeting_info}
\end{table}

\subsection{Targeting Explanations by Ad Type} 

\subsubsection{Local Ads}
\begin{figure}[b!]
\centering
\includegraphics[width=.9\linewidth]{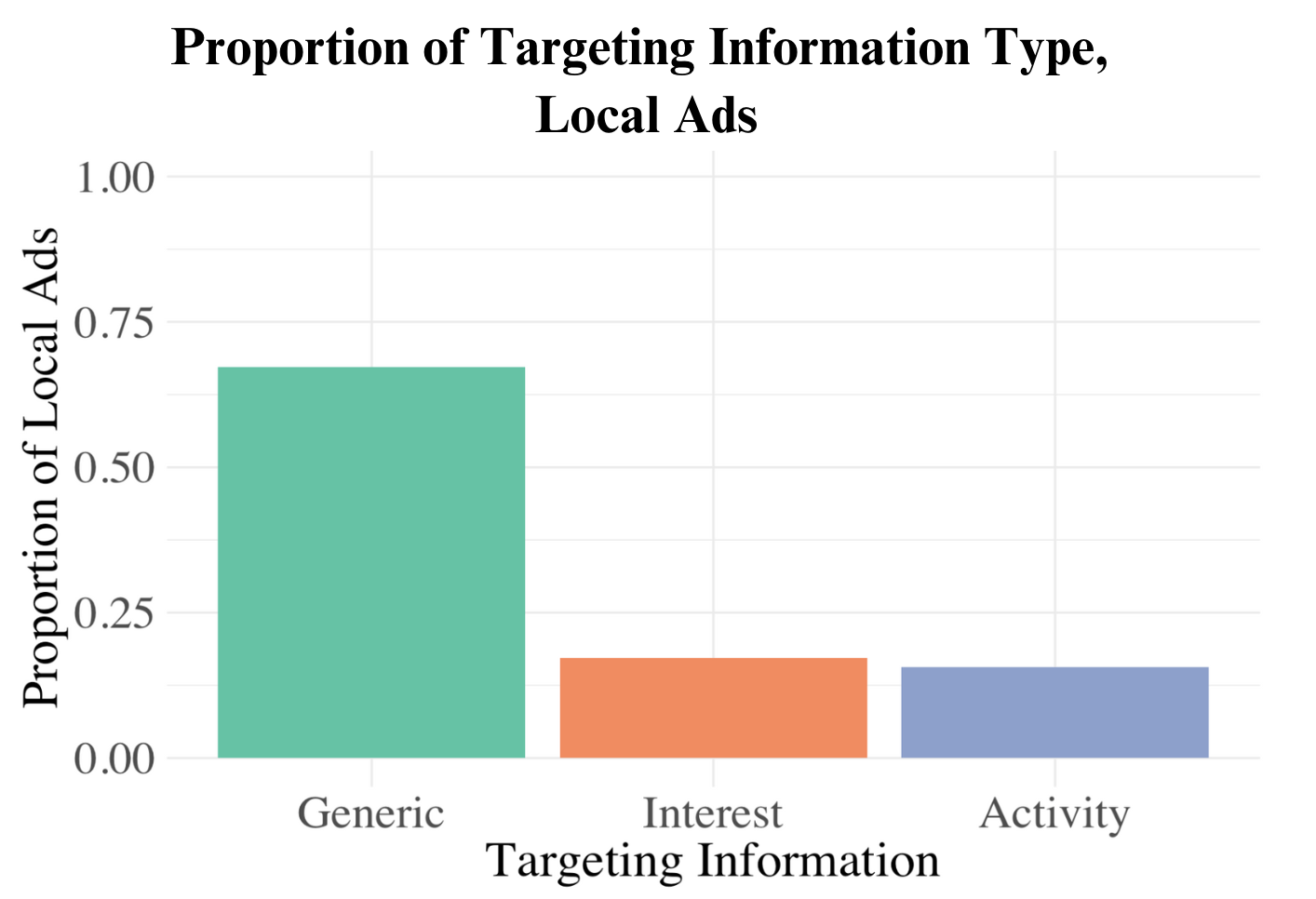}
\caption{Proportion of targeting information type for local ads.}
\label{fig:local_targeting_info}
\end{figure}

We now present our findings for local ads and their explanations (see Figure \ref{fig:local_targeting_info}). To uphold actionability and faithfulness to ad content, we expect local ads to specify either location-specific targeting or activity-related information in the explanations. We find that this is not the case -- 67\% of local ads only included generic targeting information, while 17\% listed interest-related targeting information and 16\% listed activity-related targeting information. Thus, for the vast majority of local ads, the explanations are not faithful or actionable, as we would expect local ads to either list local targeting or activity-based targeting as reasoning why it was delivered to specific users. We cannot reliably tell whether the underlying reason for generic targeting is Advantage+, but AI-mediated audience selection is one likely explanation for such a prevalence of missing local or activity information. The lack of faithfulness for local ads makes it challenging for users to understand what controls to leverage.

\subsubsection{Interest-Specific Ads} 
We now present our analysis of the targeting information attached to the 222 related ads delivered in Round 2, one week after the ``See less'' intervention. We focus on Round 2 since these ads were delivered after we would expect the intervention to be fully effective. Therefore, we expect faithful explanations to inform participants why an ad was delivered despite the control, rather than returning generic targeting explanations that do not reference the user's expressed preferences to ``See less.'' 

\begin{figure}[ht]
\centering
\includegraphics[width=\linewidth]{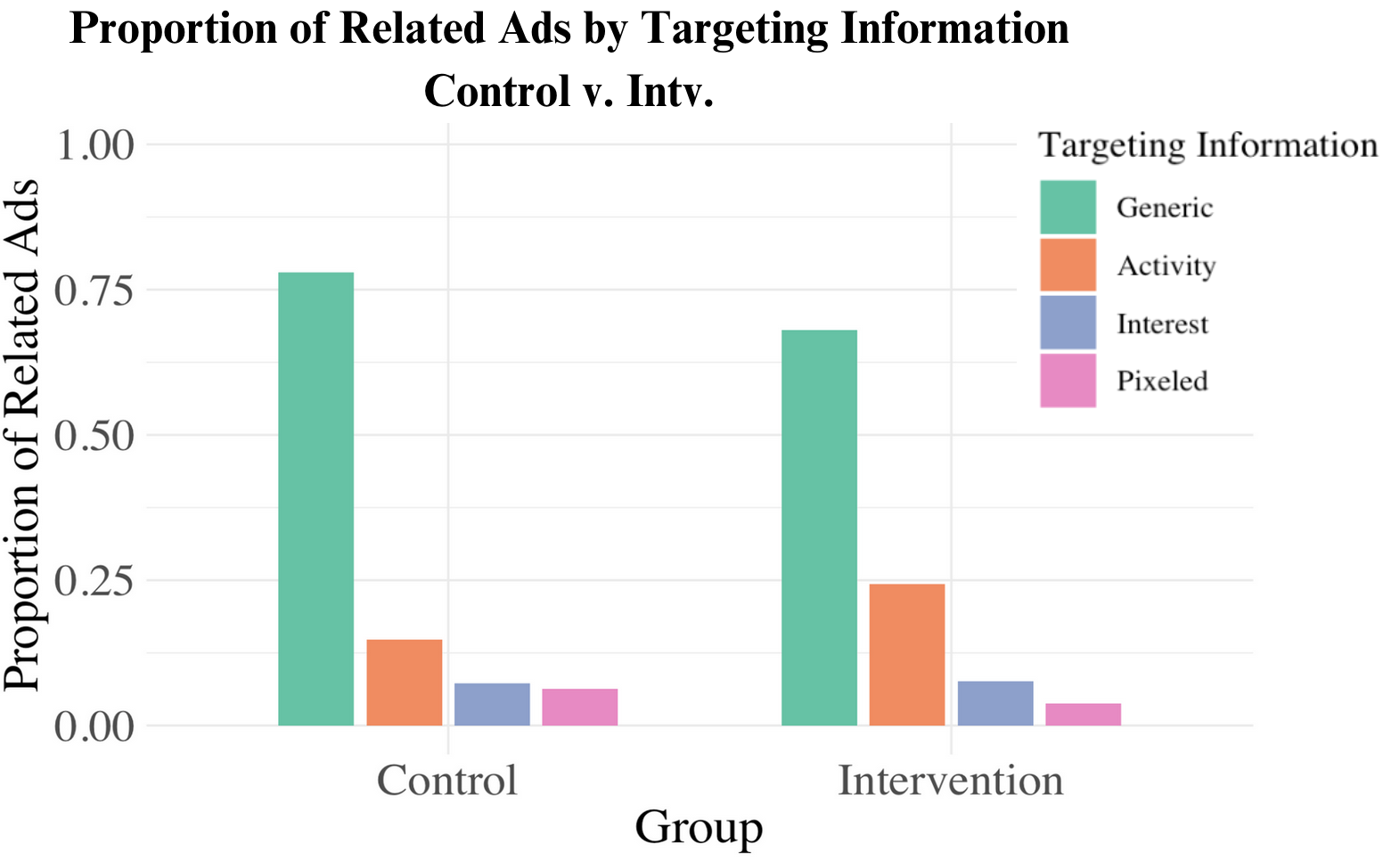}
\caption{Proportions of targeting information type for related ads, post-intervention.}
\label{fig:targeting_info_related}
\end{figure}

We calculate the proportion of generic, interest-based, and activity-based targeting information for the control and intervention groups, shown in Figure \ref{fig:targeting_info_related}. 
We also include the proportion of ads from advertisers that participants with no related ads were instructed to visit (``Pixeled'' advertisers) from Table \ref{table:sites}. 

Similar to our findings with respect to explanations for local ads, we find that the majority of targeting explanations for interest-specific ads are generic; moreover, the intervention group sees an increase in activity-based targeting explanations. Both of these findings are consistent with what one would expect from AI-mediated targeting and delivery, since AI-mediated targeting is not aligned with the interest-based ``See less'' ad control, so users are targeted despite the control, returning a majority of generic targeting explanations. 

In summary, the prevalence of generic targeting information reduces the actionability of explanations, since they do not help users understand why ads were delivered to them or express their ad preferences to change the ads they see.

\section{Discussion}
We demonstrated the ineffectiveness of Facebook's ``See less'' ad topic control and the lack of actionability in ad explanations, evidenced by the prevalence of generic ad targeting explanations. Along with \citeauthor{Chouaki_Bouzenia_Goga_Roussillon_2022}, we attribute these issues in user-facing controls and transparency tools to the recent shift from advertiser microtargeting to AI-mediated audience selection \cite{Meta_2022}.

Broadly, our findings illustrate the significant challenges in preserving transparency, actionability, and user control over complex AI systems. Previous research has highlighted the difficulty of balancing data deletion with explainable decisions in AI-mediated recommendation systems \cite{Pawelczyk_Leemann_Biega_Kasneci_2023}. As new data privacy regulations and governance protections emerge, it is crucial to investigate how AI systems can support users' rights to transparency and control while employing AI-mediated targeting mechanisms.

More specifically, our findings reflect a fundamental misalignment between Facebook's existing ad controls and explanations with their new ad delivery system. Advantage+ audience selection shows how AI can improve ad targeting while reducing the collection of user data, potentially improving privacy, but this improvement currently comes at the cost of effectiveness of user-facing control and faithfulness in the transparency of ad targeting decisions. Therefore, future research, especially from platforms, is necessary to develop new mechanisms for transparency and user control that align with the shift to AI-mediated ad targeting and delivery systems.

\subsection{Limitations \& Future Work}\label{sec:limitations}
First, our study collected a relatively low number of ads, collecting a maximum of 22 ads per participant, instead aiming to attract and retain a high number of participants. Future research could conduct a longitudinal study similar to the work done by the Panoptykon Foundation \cite{Sapieżyński_2023}, but with a larger number of participants. Additionally, due to our limited budget for recruiting participants and the difficulty of finding participants with high proportions of related ads, we asked participants to artificially change their web activity by visiting sites related to their blocked topics. The construction of surveys and collection of data could be streamlined with a browser extension hosting surveys and automatically collecting participants' ad data as in \citeauthor{Lam_Pandit_Kalicki_Gupta_Sahoo_Metaxa_2023}.

\subsection{Ethics}\label{sec:ethics}
Because of the potentially private nature of the ads users see, to minimize potential harm, we collected the minimum data necessary for our research goals and made the data sharing with us optional when possible. 
Our study and data collection and storage practices were approved by our Institutional Review Board. 
We ensured participants were fully briefed on the scope of the study prior to consenting and paid them between \$12/hr and \$20/hr, increasing payments for more time- and data-intensive surveys. 
We believe that participation in our study, such as clicking on ``Why this ad?" information did not affect users' experiences on the Facebook platform and exercising the ``See less" controls both had limited effect and was transparent to users and they could have chosen not to follow this instruction, without penalty.
We did not use any automated scraping to collect the data.

\section{Conclusion}\label{sec:conclusions}
In our large-scale study, we found that participants marking ``See less'' to \textit{Parenting} and \textit{Body Weight Control} did not see a significant decrease in the proportion of ads related to these topics over time. Furthermore, some users in correlated demographics with their blocked topics saw a higher proportion of related ads, on average, than users not in correlated demographics. 
Finally, the majority of local ads and topic-related ads delivered to users post-intervention included only generic targeting information, failing to guide users as to what controls they could further exercise to decrease the proportion of ads on these topics.
Overall, our study demonstrates that the push for advertisers to use Advantage+ audiences and the resulting AI-mediated user targeting, without the corresponding updates to the technology behind explanations and controls, leads to the prevalence of generic targeting information and the ineffectiveness of the ``See less'' ad controls. 
Our work motivates future research into transparency and control interfaces that are effective when the primary targeting process is AI- rather than advertiser- driven.

\section{Acknowledgements}\label{sec:acknow}
We thank Andrés Monroy-Hernández of the Princeton HCI Lab for his thoughtful and insightful feedback. 
This work was supported in part by funding from Princeton's McIntosh Independent Work/Senior Thesis Fund, which was used towards compensation of our survey participants and by the National Science Foundation grants
  CNS-1956435 and CNS-2344925.

\bibliography{aaai24}

\begin{thebibliography}{68}
\providecommand{\natexlab}[1]{#1}

\bibitem[{ACLU(2018)}]{ACLU_suit}
ACLU. 2018.
\newblock ACLU and Workers Take On Facebook for Gender Discrimination in Job Ads.
\newblock \url{https://www.aclu.org/press-releases/aclu-and-workers-take-facebook-gender-discrimination-job-ads}.

\bibitem[{Ali et~al.(2022)Ali, Goetzen, Mislove, Redmiles, and Sapiezynski}]{Ali_Goetzen_Mislove_Redmiles_Sapiezynski_2022}
Ali, M.; Goetzen, A.; Mislove, A.; Redmiles, E.; and Sapiezynski, P. 2022.
\newblock All Things Unequal: Measuring Disparity of Potentially Harmful Ads on {F}acebook.
\newblock In \emph{Proceedings of the 2022 Workshop on Consumer Protection}.

\bibitem[{Ali et~al.(2023)Ali, Goetzen, Mislove, Redmiles, and Sapiezynski}]{Ali_Goetzen_Mislove_Redmiles_Sapiezynski_2023}
Ali, M.; Goetzen, A.; Mislove, A.; Redmiles, E.~M.; and Sapiezynski, P. 2023.
\newblock Problematic Advertising and its Disparate Exposure on {F}acebook.
\newblock In \emph{Proceedings of the 32nd USENIX Security Symposium}, arXiv:2306.06052. arXiv.
\newblock ArXiv:2306.06052 [cs].

\bibitem[{Ali et~al.(2019)Ali, Sapiezynski, Bogen, Korolova, Mislove, and Rieke}]{Ali_Sapiezynski_Bogen_Korolova_Mislove_Rieke_2019}
Ali, M.; Sapiezynski, P.; Bogen, M.; Korolova, A.; Mislove, A.; and Rieke, A. 2019.
\newblock Discrimination through Optimization: How Facebook’s Ad Delivery Can Lead to Biased Outcomes.
\newblock \emph{Proceedings of the ACM on Human-Computer Interaction}, 3(CSCW): 1–30.

\bibitem[{Ali et~al.(2021)Ali, Sapiezynski, Korolova, Mislove, and Rieke}]{ali2021ad}
Ali, M.; Sapiezynski, P.; Korolova, A.; Mislove, A.; and Rieke, A. 2021.
\newblock Ad Delivery Algorithms: The Hidden Arbiters of Political Messaging.
\newblock In \emph{14th ACM International Conference on Web Search and Data Mining (WSDM)}.

\bibitem[{Andreou et~al.(2018)Andreou, Venkatadri, Goga, Gummadi, Loiseau, and Mislove}]{Andreou_Venkatadri_Goga_Gummadi_Loiseau_Mislove_2018}
Andreou, A.; Venkatadri, G.; Goga, O.; Gummadi, K.~P.; Loiseau, P.; and Mislove, A. 2018.
\newblock Investigating Ad Transparency Mechanisms in Social Media: A Case Study of {F}acebook’s Explanations.
\newblock In \emph{Proceedings 2018 Network and Distributed System Security Symposium}. San Diego, CA: Internet Society.
\newblock ISBN 978-1-891562-49-5.

\bibitem[{Austin(2022)}]{FacebookvsHUD}
Austin, R.~L. 2022.
\newblock Expanding Our Work on Ads Fairness.
\newblock \url{https://about.fb.com/news/2022/06/expanding-our-work-on-ads-fairness/}.

\bibitem[{Austin(2023)}]{FacebookvsHUD2023}
Austin, R.~L. 2023.
\newblock An Update on Our Ads Fairness Efforts.
\newblock \url{https://about.fb.com/news/2023/01/an-update-on-our-ads-fairness-efforts/}.

\bibitem[{Bogen et~al.(2023)Bogen, Tripathi, Timmaraju, Mashayekhi, Zeng, Roudani, Gahagan, Howard, and Leone}]{Bogen2023}
Bogen, M.; Tripathi, P.; Timmaraju, A.~S.; Mashayekhi, M.; Zeng, Q.; Roudani, R.; Gahagan, S.; Howard, A.; and Leone, I. 2023.
\newblock Toward fairness in personalized ads.

\bibitem[{Brockell(2018)}]{Brockell_2018}
Brockell, G. 2018.
\newblock Perspective | Dear tech companies, I don’t want to see pregnancy ads after my child was stillborn.
\newblock \emph{Washington Post}.

\bibitem[{Bui and Miller(2018)}]{Bui_Miller_2018}
Bui, Q.; and Miller, C.~C. 2018.
\newblock The Age That Women Have Babies: How a Gap Divides {A}merica.
\newblock \emph{The New York Times}.

\bibitem[{Caba\~{n}as, Cuevas, and Cuevas(2018)}]{Cabanas_Cuevas_Cuevas}
Caba\~{n}as, J.~G.; Cuevas, A.; and Cuevas, R. 2018.
\newblock Unveiling and quantifying facebook exploitation of sensitive personal data for advertising purposes.
\newblock In \emph{Proceedings of the 27th USENIX Conference on Security Symposium}, SEC'18, 479–495. USA: USENIX Association.
\newblock ISBN 9781931971461.

\bibitem[{Charalambides(2021)}]{prolific_demographics}
Charalambides, N. 2021.
\newblock We recently went viral on TikTok - here's what we learned.
\newblock \url{https://www.prolific.com/resources/we-recently-went-viral-on-tiktok-heres-what-we-learned}.

\bibitem[{Chouaki et~al.(2022)Chouaki, Bouzenia, Goga, and Roussillon}]{Chouaki_Bouzenia_Goga_Roussillon_2022}
Chouaki, S.; Bouzenia, I.; Goga, O.; and Roussillon, B. 2022.
\newblock Exploring the Online Micro-targeting Practices of Small, Medium, and Large Businesses.
\newblock \emph{Proceedings of the ACM on Human-Computer Interaction}, 6(CSCW2): 1–23.

\bibitem[{Chromik et~al.(2019)Chromik, Eiband, Völkel, and Buschek}]{Chromik_Eiband_Völkel_Buschek_2019}
Chromik, M.; Eiband, M.; Völkel, S.~T.; and Buschek, D. 2019.
\newblock Dark Patterns of Explainability, Transparency, and User Control for Intelligent Systems.
\newblock In \emph{Joint Proceedings of the ACM IUI 2019 Workshops}.

\bibitem[{Contreras(2022)}]{Contreras_2022}
Contreras, B. 2022.
\newblock \url{https://www.latimes.com/business/technology/story/2022-05-25/for-pregnant-women-the-internet-can-be-a-nightmare}.

\bibitem[{Datta, Tschantz, and Datta(2015)}]{Datta_Tschantz_Datta_2015}
Datta, A.; Tschantz, M.~C.; and Datta, A. 2015.
\newblock Automated Experiments on Ad Privacy Settings: A Tale of Opacity, Choice, and Discrimination.
\newblock In \emph{Proceedings on Privacy Enhancing Technologies}, arXiv:1408.6491. arXiv.
\newblock ArXiv:1408.6491 [cs].

\bibitem[{Dolin et~al.(2018)Dolin, Weinshel, Shan, Hahn, Choi, Mazurek, and Ur}]{Dolin_Weinshel_Shan_Hahn_Choi_Mazurek_Ur_2018}
Dolin, C.; Weinshel, B.; Shan, S.; Hahn, C.~M.; Choi, E.; Mazurek, M.~L.; and Ur, B. 2018.
\newblock Unpacking Perceptions of Data-Driven Inferences Underlying Online Targeting and Personalization.
\newblock In \emph{Proceedings of the 2018 CHI Conference on Human Factors in Computing Systems}, 1–12. Montreal QC Canada: ACM.
\newblock ISBN 978-1-4503-5620-6.

\bibitem[{El~fraihi et~al.(2024)El~fraihi, Amieur, Rudametkin, and Goga}]{El_Fraihi_Amieur_Rudametkin_Goga_2024}
El~fraihi, A.; Amieur, N.; Rudametkin, W.; and Goga, O. 2024.
\newblock Client-side and Server-side Tracking on Meta: Effectiveness and Accuracy.
\newblock In \emph{Proceedings on Privacy Enhancing Technologies}.

\bibitem[{Eslami et~al.(2018)Eslami, Krishna~Kumaran, Sandvig, and Karahalios}]{Eslami_Krishna_Kumaran_Sandvig_Karahalios_2018}
Eslami, M.; Krishna~Kumaran, S.~R.; Sandvig, C.; and Karahalios, K. 2018.
\newblock Communicating Algorithmic Process in Online Behavioral Advertising.
\newblock In \emph{Proceedings of the 2018 CHI Conference on Human Factors in Computing Systems}, CHI ’18, 1–13. New York, NY, USA: Association for Computing Machinery.
\newblock ISBN 978-1-4503-5620-6.

\bibitem[{{Facebook Help Center}(2024)}]{FacebookSeeLessDescription}
{Facebook Help Center}. 2024.
\newblock Choose to see less of certain ad topics while on Facebook.
\newblock \url{https://www.facebook.com/help/353660662271696}.

\bibitem[{Faizullabhoy and Korolova(2018)}]{faizullabhoy2018facebook}
Faizullabhoy, I.; and Korolova, A. 2018.
\newblock Facebook's Advertising Platform: New Attack Vectors and the Need for Interventions.
\newblock In \emph{IEEE Workshop on Technology and Consumer Protection (ConPro ’18)}.

\bibitem[{Feathers et~al.(2022)Feathers, Fondrie-Teitler, Waller, and Mattu}]{Feathers_Fondrie_Mattu_2022}
Feathers, T.; Fondrie-Teitler, S.; Waller, A.; and Mattu, S. 2022.
\newblock {F}acebook Is Receiving Sensitive Medical Information from Hospital Websites – {T}he {M}arkup.
\newblock \url{https://themarkup.org/pixel-hunt/2022/06/16/facebook-is-receiving-sensitive-medical-information-from-hospital-websites}.

\bibitem[{Gak, Olojo, and Salehi(2022)}]{Gak_Olojo_Salehi_2022}
Gak, L.; Olojo, S.; and Salehi, N. 2022.
\newblock The Distressing Ads That Persist: Uncovering The Harms of Targeted Weight-Loss Ads Among Users with Histories of Disordered Eating.
\newblock \emph{Proceedings of the ACM on Human-Computer Interaction}, 6(CSCW2): 1–23.

\bibitem[{Habib et~al.(2020)Habib, Pearman, Wang, Zou, Acquisti, Cranor, Sadeh, and Schaub}]{Habib_Pearman_Wang_Zou_Acquisti_Cranor_Sadeh_Schaub_2020}
Habib, H.; Pearman, S.; Wang, J.; Zou, Y.; Acquisti, A.; Cranor, L.~F.; Sadeh, N.; and Schaub, F. 2020.
\newblock “{I}t’s a scavenger hunt”: Usability of Websites’ Opt-Out and Data Deletion Choices.
\newblock In \emph{Proceedings of the 2020 CHI Conference on Human Factors in Computing Systems}, CHI ’20, 1–12. New York, NY, USA: Association for Computing Machinery.
\newblock ISBN 978-1-4503-6708-0.

\bibitem[{Habib et~al.(2022)Habib, Pearman, Young, Saxena, Zhang, and Cranor}]{Habib_Pearman_Young_Saxena_Zhang_Cranor_2022}
Habib, H.; Pearman, S.; Young, E.; Saxena, I.; Zhang, R.; and Cranor, L.~F. 2022.
\newblock Identifying User Needs for Advertising Controls on {F}acebook.
\newblock \emph{Proceedings of the ACM on Human-Computer Interaction}, 6(CSCW1): 1–42.

\bibitem[{Hsu et~al.(2020)Hsu, Vaccaro, Yue, Rickman, and Karahalios}]{Hsu_Vaccaro_Yue_Rickman_Karahalios_2020}
Hsu, S.; Vaccaro, K.; Yue, Y.; Rickman, A.; and Karahalios, K. 2020.
\newblock Awareness, Navigation, and Use of Feed Control Settings Online.
\newblock In \emph{Proceedings of the 2020 CHI Conference on Human Factors in Computing Systems}, CHI ’20, 1–13. New York, NY, USA: Association for Computing Machinery.
\newblock ISBN 978-1-4503-6708-0.

\bibitem[{Imana, Korolova, and Heidemann(2021)}]{imana2021auditing}
Imana, B.; Korolova, A.; and Heidemann, J. 2021.
\newblock Auditing for Discrimination in Algorithms Delivering Job Ads.
\newblock In \emph{The Web Conference (WWW)}.

\bibitem[{Imana, Korolova, and Heidemann(2024)}]{Imana_Korolova_Heidemann_2024}
Imana, B.; Korolova, A.; and Heidemann, J. 2024.
\newblock Auditing for Racial Discrimination in the Delivery of Education Ads.
\newblock In \emph{The 2024 ACM Conference on Fairness, Accountability, and Transparency}, 2348–2361.
\newblock ArXiv:2406.00591 [cs].

\bibitem[{Korolova(2011)}]{korolova2011privacy}
Korolova, A. 2011.
\newblock Privacy Violations Using Microtargeted Ads: A Case Study.
\newblock \emph{Journal of Privacy and Confidentiality}, 3(1): 27--49.

\bibitem[{Lam et~al.(2023)Lam, Pandit, Kalicki, Gupta, Sahoo, and {M}etaxa}]{Lam_Pandit_Kalicki_Gupta_Sahoo_Metaxa_2023}
Lam, M.~S.; Pandit, A.; Kalicki, C.~H.; Gupta, R.; Sahoo, P.; and {M}etaxa, D. 2023.
\newblock Sociotechnical Audits: Broadening the Algorithm Auditing Lens to Investigate Targeted Advertising.
\newblock \emph{Proceedings of the ACM on Human-Computer Interaction}, 7(CSCW2): 360:1--360:37.

\bibitem[{Lee et~al.(2023)Lee, Logas, Yang, Li, Barbosa, Wang, and Das}]{Lee_Logas_Yang_Li_Barbosa_Wang_Das_2023}
Lee, H.-P.~H.; Logas, J.; Yang, S.; Li, Z.; Barbosa, N.; Wang, Y.; and Das, S. 2023.
\newblock When and Why Do People Want Ad Targeting Explanations? {E}vidence from a Four-Week, Mixed-Methods Field Study.
\newblock In \emph{2023 IEEE Symposium on Security and Privacy (SP)}, 2903–2920. San Francisco, CA, USA: IEEE.
\newblock ISBN 978-1-66549-336-9.

\bibitem[{Leon et~al.(2012)Leon, Ur, Shay, Wang, Balebako, and Cranor}]{Leon_Ur_Shay_Wang_Balebako_Cranor_2012}
Leon, P.; Ur, B.; Shay, R.; Wang, Y.; Balebako, R.; and Cranor, L. 2012.
\newblock Why Johnny can’t opt out: A usability evaluation of tools to limit online behavioral advertising.
\newblock In \emph{Proceedings of the SIGCHI Conference on Human Factors in Computing Systems}, 589–598. Austin Texas USA: ACM.
\newblock ISBN 978-1-4503-1015-4.

\bibitem[{Lipton(2018)}]{Lipton_2018}
Lipton, Z.~C. 2018.
\newblock The Mythos of Model Interpretability: In machine learning, the concept of interpretability is both important and slippery.
\newblock \emph{Queue}, 16(3): 31–57.

\bibitem[{Mann and Whitney(1947)}]{Mann_Whitney_1947}
Mann, H.~B.; and Whitney, D.~R. 1947.
\newblock On a Test of Whether one of Two Random Variables is Stochastically Larger than the Other.
\newblock \emph{The Annals of Mathematical Statistics}, 18(1): 50–60.

\bibitem[{Martin et~al.(2019)Martin, Herrick, Sarafrazi, and Ogden}]{Martin_Herrick_Sarafrazi_Ogden_2019}
Martin, C.; Herrick, K.; Sarafrazi, N.; and Ogden, C. 2019.
\newblock Attempts to Lose Weight Among Adults in the {U}nited {S}tates, 2013–2016.
\newblock \url{https://www.cdc.gov/nchs/products/databriefs/db313.htm}.

\bibitem[{Mattu and Sankin(2020)}]{Mattu_Sankin_2020}
Mattu, S.; and Sankin, A. 2020.
\newblock How We Built a Real-time Privacy Inspector – {T}he {M}arkup.
\newblock \url{https://themarkup.org/blacklight/2020/09/22/how-we-built-a-real-time-privacy-inspector}.

\bibitem[{Meta(2019)}]{waist}
Meta. 2019.
\newblock Why Am I Seeing This? We Have an Answer for You.
\newblock \url{https://about.fb.com/news/2019/03/why-am-i-seeing-this/}.

\bibitem[{Meta(2020)}]{Facebook_target_2020}
Meta. 2020.
\newblock Simplifying Targeting Categories.
\newblock \url{https://www.facebook.com/business/news/update-to-facebook-ads-targeting-categories}.

\bibitem[{Meta(2021)}]{Facebook_target_2022}
Meta. 2021.
\newblock Removing Certain Ad Targeting Options and Expanding Our Ad Controls.
\newblock \url{https://www.facebook.com/business/news/removing-certain-ad-targeting-options-and-expanding-our-ad-controls}.

\bibitem[{Meta(2022{\natexlab{a}})}]{Meta_2022}
Meta. 2022{\natexlab{a}}.
\newblock Introducing New Automation Tools to Increase Sales and Drive Growth.
\newblock \url{https://about.fb.com/news/2022/08/introducing-new-automation-tools-to-increase-sales-and-drive-growth/}.
\newblock Accessed: 2024-05-04.

\bibitem[{Meta(2022{\natexlab{b}})}]{Facebook_target_2022b}
Meta. 2022{\natexlab{b}}.
\newblock Preparing for Upcoming Removal of Certain Ad Targeting Options.
\newblock \url{https://www.facebook.com/government-nonprofits/blog/preparing-for-upcoming-removal-of-certain-ad-targeting-options}.

\bibitem[{Meta(2023)}]{Meta_Transparency_2023}
Meta. 2023.
\newblock Increasing Our Ads Transparency.
\newblock \url{https://about.fb.com/news/2023/02/increasing-our-ads-transparency/}.

\bibitem[{Meta(2024{\natexlab{a}})}]{Facebook_Advantage+_2024}
Meta. 2024{\natexlab{a}}.
\newblock About Advantage+ audience.
\newblock \url{https://www.facebook.com/business/help/273363992030035}.

\bibitem[{Meta(2024{\natexlab{b}})}]{2024_q1_ad_revenue}
Meta. 2024{\natexlab{b}}.
\newblock Meta Reports First Quarter 2024 Results.
\newblock \url{https://investor.fb.com/investor-news/press-release-details/2024/Meta-Reports-First-Quarter-2024-Results/default.aspx}.

\bibitem[{Meta(2024{\natexlab{c}})}]{Facebook_target_2024}
Meta. 2024{\natexlab{c}}.
\newblock Updates to detailed targeting.
\newblock \url{https://www.facebook.com/business/help/458835214668072}.

\bibitem[{Nagaraj~Rao and Korolova(2023)}]{Nagaraj_Rao_Korolova_2023}
Nagaraj~Rao, V.; and Korolova, A. 2023.
\newblock Discrimination through Image Selection by Job Advertisers on {F}acebook.
\newblock In \emph{2023 ACM Conference on Fairness, Accountability, and Transparency}, 1772–1788. Chicago IL USA: ACM.
\newblock ISBN 9798400701924.

\bibitem[{Pawelczyk et~al.(2023)Pawelczyk, Leemann, Biega, and Kasneci}]{Pawelczyk_Leemann_Biega_Kasneci_2023}
Pawelczyk, M.; Leemann, T.; Biega, A.; and Kasneci, G. 2023.
\newblock On the Trade-Off between Actionable Explanations and the Right to be Forgotten.
\newblock In \emph{International Conference on Learning Representations}, arXiv:2208.14137. arXiv.
\newblock ArXiv:2208.14137 [cs].

\bibitem[{Pesenti(2021)}]{Pesenti_2021}
Pesenti, J. 2021.
\newblock {F}acebook’s five pillars of Responsible {AI}.
\newblock \url{https://ai.meta.com/blog/facebooks-five-pillars-of-responsible-ai/}.

\bibitem[{Poursabzi-Sangdeh et~al.(2021)Poursabzi-Sangdeh, Goldstein, Hofman, Wortman~Vaughan, and Wallach}]{Poursabzi_Sangdeh_Goldstein_2021}
Poursabzi-Sangdeh, F.; Goldstein, D.~G.; Hofman, J.~M.; Wortman~Vaughan, J.~W.; and Wallach, H. 2021.
\newblock Manipulating and Measuring Model Interpretability.
\newblock In \emph{Proceedings of the 2021 CHI Conference on Human Factors in Computing Systems}, CHI ’21, 1–52. New York, NY, USA: Association for Computing Machinery.
\newblock ISBN 978-1-4503-8096-6.

\bibitem[{Prolific(2024)}]{prolific}
Prolific. 2024.
\newblock Easily find vetted research participants and AI taskers at scale.
\newblock \url{https://www.prolific.com/}.

\bibitem[{Rader and Gray(2015)}]{Rader_Gray_2015}
Rader, E.; and Gray, R. 2015.
\newblock Understanding User Beliefs About Algorithmic Curation in the Facebook News Feed.
\newblock In \emph{Proceedings of the 33rd Annual ACM Conference on Human Factors in Computing Systems}, CHI ’15, 173–182. New York, NY, USA: Association for Computing Machinery.
\newblock ISBN 978-1-4503-3145-6.

\bibitem[{Ribeiro et~al.(2019)Ribeiro, Saha, Babaei, Henrique, Messias, Benevenuto, Goga, Gummadi, and Redmiles}]{ribeiro2019microtargeting}
Ribeiro, F.~N.; Saha, K.; Babaei, M.; Henrique, L.; Messias, J.; Benevenuto, F.; Goga, O.; Gummadi, K.~P.; and Redmiles, E.~M. 2019.
\newblock On microtargeting socially divisive ads: A case study of russia-linked ad campaigns on facebook.
\newblock In \emph{Proceedings of the conference on fairness, accountability, and transparency}, 140--149.

\bibitem[{Sampson, Encarnacion, and {M}etaxa(2023)}]{Sampson_Encarnacion_Metaxa_2023}
Sampson, P.; Encarnacion, R.; and {M}etaxa, D. 2023.
\newblock Representation, Self-Determination, and Refusal: Queer People’s Experiences with Targeted Advertising.
\newblock In \emph{2023 ACM Conference on Fairness, Accountability, and Transparency}, 1711–1722. Chicago IL USA: ACM.
\newblock ISBN 9798400701924.

\bibitem[{Sapiezynski et~al.(2022)Sapiezynski, Ghosh, Kaplan, Rieke, and Mislove}]{Sapiezynski_Ghosh_Kaplan_Rieke_Mislove_2022}
Sapiezynski, P.; Ghosh, A.; Kaplan, L.; Rieke, A.; and Mislove, A. 2022.
\newblock Algorithms that “Don’t See Color”: Comparing Biases in Lookalike and Special Ad Audiences.
\newblock In \emph{Proceedings of the 2022 AAAI/ACM Conference on AI, Ethics, and Society}, 609–616.
\newblock ArXiv:1912.07579 [cs].

\bibitem[{Sapiezynski et~al.(2024)Sapiezynski, Kaplan, Korolova, and Mislove}]{Proxies2024}
Sapiezynski, P.; Kaplan, L.; Korolova, A.; and Mislove, A. 2024.
\newblock On the Use of Proxies in Political Ad Targeting.
\newblock In \emph{Proceedings of the ACM on Human-Computer Interaction}.

\bibitem[{Sapieżyński(2023)}]{Sapieżyński_2023}
Sapieżyński, P. 2023.
\newblock Algorithms of Trauma 2. {H}ow {F}acebook Feeds on Your Fears | {F}undacja {P}anoptykon.
\newblock \url{https://panoptykon.org/algorithms-of-trauma-2-how-facebook-feeds-on-your-fears}.

\bibitem[{Shane(2017)}]{Shane_2017}
Shane, S. 2017.
\newblock These Are the Ads {R}ussia Bought on {F}acebook in 2016.
\newblock \url{https://www.nytimes.com/2017/11/01/us/politics/russia-2016-election-facebook.html}.

\bibitem[{Smith-Renner et~al.(2020)Smith-Renner, Fan, Birchfield, Wu, Boyd-Graber, Weld, and Findlater}]{Smith-Renner_Fan_Birchfield_Wu_Boyd-Graber_Weld_Findlater_2020}
Smith-Renner, A.; Fan, R.; Birchfield, M.; Wu, T.; Boyd-Graber, J.; Weld, D.~S.; and Findlater, L. 2020.
\newblock No Explainability without Accountability: An Empirical Study of Explanations and Feedback in Interactive ML.
\newblock In \emph{Proceedings of the 2020 CHI Conference on Human Factors in Computing Systems}, 1–13. Honolulu HI USA: ACM.
\newblock ISBN 978-1-4503-6708-0.

\bibitem[{Speicher et~al.(2018)Speicher, Ali, Venkatadri, Ribeiro, Arvanitakis, Benevenuto, Gummadi, Loiseau, and Mislove}]{Speicher_Ali_Venkatadri_Ribeiro_Arvanitakis_Benevenuto_Gummadi_Loiseau_Mislove}
Speicher, T.; Ali, M.; Venkatadri, G.; Ribeiro, F.~N.; Arvanitakis, G.; Benevenuto, F.; Gummadi, K.~P.; Loiseau, P.; and Mislove, A. 2018.
\newblock Potential for Discrimination in Online Targeted Advertising.
\newblock \emph{Proceedings of Machine Learning Research}.

\bibitem[{{The US Department of Justice}(2022)}]{FacebookvsHUD2}
{The US Department of Justice}. 2022.
\newblock {Justice Department Secures Groundbreaking Settlement Agreement with Meta Platforms, Formerly Known as Facebook, to Resolve Allegations of Discriminatory Advertising}.
\newblock \url{https://www.justice.gov/opa/pr/justice-department-secures-groundbreaking-settlement-agreement-meta-platforms-formerly-known}.

\bibitem[{Timmaraju et~al.(2023)Timmaraju, Mashayekhi, Chen, Zeng, Fettes, Cheung, Xiao, Kannadasan, Tripathi, Gahagan, Bogen, and Roudani}]{Timmaraju_Mashayekhi_Chen_Zeng_Fettes_Cheung_Xiao_Kannadasan_Tripathi_Gahagan_2023}
Timmaraju, A.~S.; Mashayekhi, M.; Chen, M.; Zeng, Q.; Fettes, Q.; Cheung, W.; Xiao, Y.; Kannadasan, M.~R.; Tripathi, P.; Gahagan, S.; Bogen, M.; and Roudani, R. 2023.
\newblock Towards Fairness in Personalized Ads Using Impression Variance Aware Reinforcement Learning.
\newblock In \emph{Proceedings of the 29th ACM SIGKDD Conference on Knowledge Discovery and Data Mining}, 4937–4947.
\newblock ArXiv:2306.03293 [cs].

\bibitem[{Venkatadri et~al.(2018)Venkatadri, Andreou, Liu, Mislove, Gummadi, Loiseau, and Goga}]{Venkatadri_Andreou_Liu_Mislove_Gummadi_Loiseau_Goga_2018}
Venkatadri, G.; Andreou, A.; Liu, Y.; Mislove, A.; Gummadi, K.~P.; Loiseau, P.; and Goga, O. 2018.
\newblock Privacy Risks with {F}acebook’s {PII}-Based Targeting: Auditing a Data Broker’s Advertising Interface.
\newblock In \emph{2018 IEEE Symposium on Security and Privacy (SP)}, 89–107. San Francisco, CA: IEEE.
\newblock ISBN 978-1-5386-4353-2.

\bibitem[{Wei et~al.(2020)Wei, Stamos, Veys, Reitinger, Goodman, Herman, Filipczuk, Weinshel, Mazurek, and Ur}]{Wei_Stamos_Veys_Reitinger_Goodman_Herman_Filipczuk_Weinshel_Mazurek_Ur_2020}
Wei, M.; Stamos, M.; Veys, S.; Reitinger, N.; Goodman, J.; Herman, M.; Filipczuk, D.; Weinshel, B.; Mazurek, M.~L.; and Ur, B. 2020.
\newblock What twitter knows: characterizing ad targeting practices, user perceptions, and ad explanations through users' own twitter data.
\newblock In \emph{Proceedings of the 29th USENIX Conference on Security Symposium}, SEC'20. USA: USENIX Association.
\newblock ISBN 978-1-939133-17-5.

\bibitem[{Weintraub(2019)}]{weintraub2019microtargeting}
Weintraub, E.~L. 2019.
\newblock Don't abolish political ads on social media. Stop microtargeting.
\newblock \url{https://www.washingtonpost.com/opinions/2019/11/01/dont-abolish-political-ads-social-media-stop-microtargeting/}.

\bibitem[{Wu et~al.(2023)Wu, Bice, Edwards, and Das}]{Wu_Bice_Edwards_Das_2023}
Wu, Y.; Bice, S.; Edwards, W.~K.; and Das, S. 2023.
\newblock The Slow Violence of Surveillance Capitalism: How Online Behavioral Advertising Harms People.
\newblock In \emph{Proceedings of the 2023 ACM Conference on Fairness, Accountability, and Transparency}, FAccT ’23, 1826–1837. New York, NY, USA: Association for Computing Machinery.
\newblock ISBN 9798400701924.

\bibitem[{Yuan et~al.(2023)Yuan, Bi, Lin, and Tseng}]{Yuan_Bi_Lin_Tseng_2023}
Yuan, C. W.~T.; Bi, N.; Lin, Y.-F.; and Tseng, Y.-H. 2023.
\newblock Contextualizing User Perceptions about Biases for Human-Centered Explainable Artificial Intelligence.
\newblock In \emph{Proceedings of the 2023 CHI Conference on Human Factors in Computing Systems}, CHI ’23, 1–15. New York, NY, USA: Association for Computing Machinery.
\newblock ISBN 978-1-4503-9421-5.

\bibitem[{Zeng, Kohno, and Roesner(2021)}]{Zeng_Kohno_Roesner_2021}
Zeng, E.; Kohno, T.; and Roesner, F. 2021.
\newblock What Makes a “Bad” Ad? {U}ser Perceptions of Problematic Online Advertising.
\newblock In \emph{Proceedings of the 2021 CHI Conference on Human Factors in Computing Systems}, 1–24. Yokohama Japan: ACM.
\newblock ISBN 978-1-4503-8096-6.

\end{thebibliography}

\end{document}